\documentclass[%
 dvips,twocolumn,endfloats*,
 ,secnumarabic%
,amssymb, amsmath,nobibnotes, aps,prb]{revtex4}
\usepackage{bm}%
\usepackage{graphicx}

\newcommand{\CO}{CO$_2\:$}
\newcommand{\misura}[1]{\, \mathrm{#1}}     
\newcommand{\samarion}{\mbox{SrB$_4$O$_7$:Sm$^{2+} \:$}}
\newcommand{\til}{~}

\begin{document}
\title{Melting curve and fluid equation of state of carbon dioxide at high pressure and high temperature}%

\author{Valentina M. Giordano}%
\email{valentina.giordano@esrf.fr}

\affiliation{IMPMC, Physique des Milieux Denses, CNRS UMR 7590, Universit\'e Pierre et Marie Curie, 140 rue
Lourmel, 75015 Paris, France,} \affiliation{LENS, Polo Scientifico, Sesto Fiorentino (FI), Italy}
\author{Fr\'ed\'eric Datchi}
\email{datchi@impmc.jussieu.fr} \affiliation{IMPMC, Physique des Milieux Denses, CNRS UMR 7590, Universit\'e
Pierre et Marie Curie, 140 rue Lourmel, 75015 Paris, France}

\author{Agn\`es Dewaele}
\affiliation{DIF/D\'epartement de Physique Th\'eorique et Appliqu\'ee, CEA, Bo\^ite Postale 12, 91680
Bruy\`eres-le-Ch\^atel, France}

\date{\today}%
\begin{abstract}
The melting curve and fluid equation of state of carbon dioxide have been determined under high pressure in a
resistively-heated diamond anvil cell. The melting line was determined from room temperature up to
$11.1\pm0.1$~GPa and $800\pm5$~K by visual observation of the solid-fluid equilibrium and in-situ measurements of
pressure and temperature. Raman spectroscopy was used to identify the solid phase in equilibrium with the melt,
showing that solid I is the stable phase along the melting curve in the probed range. Interferometric and
Brillouin scattering experiments were conducted to determine the refractive index and sound velocity of the fluid
phase. A dispersion of the sound velocity between ultrasonic and Brillouin frequencies is evidenced and could be
reproduced by postulating the presence of a thermal relaxation process. The Brillouin sound velocities were then
transformed to thermodynamic values in order to calculate the equation of state of fluid \CO. An analytic
formulation of the density with respect to pressure and temperature is proposed, suitable in the $P-T$ range
0.1-8~GPa and 300-700~K and  accurate within 2\%. Our results show that the fluid above 500 K is less
compressible than predicted from various phenomenological  models.
\end{abstract}
\pacs{} \keywords{} \maketitle

\section{Introduction}

Knowledge of the thermodynamic properties of simple molecular fluids like water and carbon dioxide at high
density is important in a large number of scientific and technological domains, such as planetary sciences or
material synthesis in supercritical conditions. On the other hand, the investigation of simple fluids at high
pressure has led  to important discoveries in the recent past, like the first-order phase transition found in
phosporus\cite{Katayama}. Such transitions can be detected by kinks in the melting curve or discontinuities in
the fluid equation of state, provided these are measured with high accuracy.

\begin{figure}
\centering
\includegraphics[width=8.5cm]{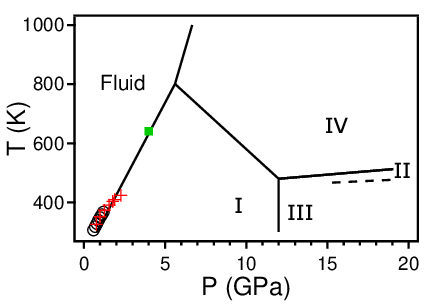}
\caption{}
\label{phasedia}
\end{figure}

The solid phase diagram of carbon dioxide has been intensively studied in the past 15 years.
Fig.\til\ref{phasedia} shows the phase diagram as presently determined for $P <
20\misura{GPa}$\til\cite{IotYoo2001,GorGio2004}. At room temperature and 0.5 GPa, \CO crystallizes into a typical molecular
solid, the cubic phase I \cite{Oli1982}, which transforms to the  orthorhombic  phase III for $P >
10\misura{GPa}$ \cite{Hanson1989,AokYam1993}. For $P > 12\misura{GPa}$ and high temperature two other molecular
phases exist: phase II for $T > 470\misura{K}$ and phase IV for $T >
500\misura{K}$\til\cite{IotYoo2001,Yoo2001,GorGio2004}.  Since both of these phases can be quenched to room
temperature without reverting to phase III, it has been proposed that phase III is actually metastable. Finally,
at higher pressures and temperatures, carbon dioxide transforms to a polymeric structure (phase V) with four-fold
coordinated carbon atoms\til\cite{IotYooCyn1999,YooCyn1999,IotYooppb2001,SanLin2004}.

In contrast with the solid phases, there have been very few studies on the high pressure fluid. To begin with,
the limit between the fluid and solid domains, i.e. the melting curve, is still barely known. The one reported in
Fig.\til\ref{phasedia} is based on data up to 366$\misura{K}$ and 1.17 GPa published by
Bridgman\til\cite{Bri1914} in 1914 and one single point measured by Iota and Yoo\til\cite{IotYoo2001} at
$4\misura{GPa}$ and $640\misura{K}$. The latter is actually in strong disagreement with the extrapolation of the
melting curve reported by Grace and Kennedy\til\cite{GraKen1967}, which gives a melting temperature lower by 130
K at this pressure. It is clear then that new and extended measurements are needed.

There is also a lack of experimental data on the fluid equation of state at high pressures. PVT measurements in static high pressure
experiments have so far been limited to 0.8 GPa  and 980 K\cite{Tsiklis1971,Shmonov1974}, while shock-wave
experiments have given a few points along the Hugoniot curve up to 70 GPa and estimated temperatures of a few
thousand K\cite{Nellis1991,Schott1991,Zubarev1962}. There is thus a gap between these two sets of data that
remains to be bridged. Attempts have been made to derive phenomenological equations of state based on various
formulations, but their accuracy in the domain where experimental data are absent is unknown. The most
sophisticated of these formulations has been proposed by Span and Wagner (SWEOS)\til\cite{SpaWag1996}, whose
validity range was estimated to extend from the triple point ($P=0.518\misura{MPa}$, $T=216.59\misura{K}$) to
$0.8\misura{GPa}$ and $1100\misura{K}$. This is based on a phenomenological form of the Helmoltz free energy with
42 free parameters fitted to a large bank of thermodynamic experimental data available in the low-pressure range.
Although very accurate within its validity range, its adequacy in the extrapolated range is questionable, as
shows the comparison between the experimental Hugoniot curve and the predicted one\cite{SpaWag1996}. Among the
other available models, the Pitzer-Sterner equation of state (PSEOS)\cite{PitSte1994} has been considered a good
one at high pressure because it was constrained by the shock wave measurements.

We present here the first study of fluid \CO at static pressures above 1$\misura{GPa}$. First, we measured the
melting curve up to 800$\misura{K}$ in order to determine the fluid stability domain.  Using Brillouin scattering
and interferometric measurements, we then determined the sound velocity and refractive index of fluid \CO up to
700$\misura{K}$ and  8$\misura{GPa}$. From the measured sound velocity, we derived the fluid equation of state in
the same $P-T$ range, which extends the previously covered pressure range by a factor of 10.

This paper is organized as follows. In a first section, we describe the experimental details of our experiments.
Then in section III we present the results obtained for the melting curve, the refractive index and sound
velocity and compare them with available literature data. In section IV, the method used to derive the equation
of state is given along with the results. Finally, the conclusion summarizes our main findings.

\section{Experimental details}

The present experiments were conducted in a membrane diamond-anvil cell (mDAC) made of high-temperature resistant
alloy (PER72 from Aubert et Duval). CO$_2$ (99.99 \% purity) samples were loaded by condensing the gas at
273$\misura{K}$ and 35$\misura{bars}$ in a high-pressure vessel. A golden ring about 10~$\mu$m large separated
the sample from the gasket material (rhenium) in order to avoid any possible chemical reaction at high
temperature. A ruby ball, a small amount of SrB$_4$O$_7$:Sm$^{2+}$ and a 15 $\mu$m-size cubic BN crystal were
loaded with the sample. Pressure was primarily determined from the $^5D_0-^7F_0$ luminescence line of
SrB$_4$O$_7$:Sm$^{2+}$ and cross-checked with the one obtained from the Raman shift of the TO mode of c-BN, using
the pressure scales reported in Refs.\til[\onlinecite{DatLeT1997,MaoXu1986,DatCan2004}]. The estimated
uncertainty on pressure was about 0.02~GPa at 300~K, 0.05~GPa at 500~K and 0.1$\misura{GPa}$ at 700 K.

The whole mDAC was heated by means of a ring-shaped resistive heater enveloping the cell, whose temperature can
be regulated within 1 K using an electronic module. Heating was done in air or in a reducing atmosphere (mixture
of Ar and 2\%H$_2$) obtained by flushing the gas directly into the cell. Because the cell is globally heated,
this setup provides minimal temperature gradients across the cell, as found from numerous previous experiments
performed in the same temperature range \cite{DatLeT1997,DatLou2000,DatCan2004}: the maximum difference observed
between the sample and other parts of the cell is 5 K. A K-type thermocouple was placed in contact with the
sides of one of the diamond anvils and fixed in place by high temperature ceramic cement.  An independent and
\textit{in-situ} measure of temperature was provided by using the ruby as a thermometer, as described in
Ref.\til[\onlinecite{DatLeT1997}]. This temperature agreed with that of the thermocouple within $\pm5$ K up to
600 K, as typically observed for this setup. At higher temperatures the ruby $R$-doublet is very broad and no
more resolved, so that this temperature measurement becomes increasingly inaccurate. We thus relied on the
temperature given by the thermocouple. The reliability of the temperature reading is supported by the very good
reproducibility of the present results between experiments performed on different samples, and by the continuous
aspect of the melting line as shown below. On this basis, we can confidently claim that the error on temperature
is within $\pm5\misura{K}$.

The melting curve was determined by visually monitoring the solid--fluid equilibrium. The sample was brought to
its room temperature melting pressure (0.55$\misura{GPa}$) where a single crystal was grown by finely tuning the
load and kept in equilibrium with its melt. The sample was then slowly heated and compressed in order to maintain
the coexistence of the crystal and the fluid, as shown in the inset of Fig.~2. The difference in refractive index
between the fluid and solid phases remained large enough to observe a good contrast between the two phases up to
our highest pressure. A melting point was defined by the measurement of pressure and temperature when equilibrium
was observed and stable in $P$ and $T$.  This method allows a fine sampling of the melting curve and precludes
any error that could originate from metastabilities such as under-cooling or over-pressurization.

The coexistent solid phase was identified by Raman spectroscopy all along the melting curve. Raman spectra  were
collected in a back-scattering geometry using the 514.53$\misura{nm}$ line from an Ar$^{+}$ laser and a T64000
Raman spectrometer (Jobin-Yvon-Horiba, $f=0.64$ m, 1800 grooves/mm grating, 100 $\mu$m entrance slits) equipped
with a $\ell$N$_2$-cooled CCD array detector.

The refractive index of the fluid was measured by an interferometric technique described in detail in Refs.
[\onlinecite{LeTLou1989}] and [\onlinecite{Dewaele2003}]. Briefly, this method involves measurements of the
interference patterns obtained by illuminating the sample with parallel white light ($500<\lambda<700$~nm) on one
hand and monochromatic light ($\lambda=632.8$~nm, Fabry-Perot rings) on the other hand. By combining the two
interference patterns, we were able to determine both thickness and refractive index $n$ of the sample. The
relative uncertainty on the refractive index measured by this method is approximately $5\times 10^{-3}$.

Brillouin scattering experiments were performed using a 6-pass tandem Fabry-Perot interferometer from JRS
Scientific \cite{Sander}. The scattered light from a Ar$^{+}$ laser excitation was collected in a back-scattering
geometry. In this case, the Brillouin frequency shift is given by:
\begin{equation}
\Delta\sigma=2nv_s/\lambda_0c_0 \label{BrillouinShift}
\end{equation}
where $v_s$, $\lambda_0=514.53$~mn and $c_0$ are respectively the sound velocity, the excitation wavelength and
the speed of light in vacuum. The frequency shift was determined with an accuracy of $2\times 10^{-3}$~cm$^{-1}$.

\section{Results and discussion}

\subsection{The melting curve}

\begin{figure}
\centering
\includegraphics[width=8.5cm]{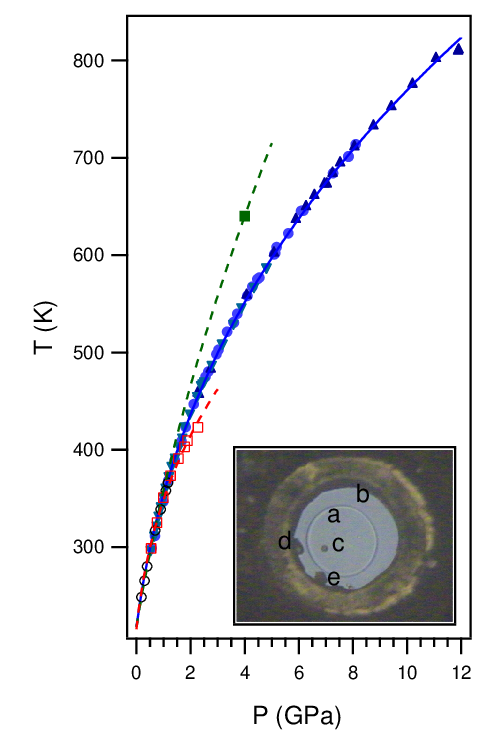}
\caption{}
\label{melting}
\end{figure}

The melting curve of \CO was determined in three separate experiments covering the temperature range 300-800 K
and pressure range 0.55-11.1 GPa. The measured melting points are shown in Fig.\til\ref{melting} along with the
available literature data. The different runs showed very good reproducibility, within the uncertainty of our
pressure and temperature readings. The large number of collected data constrain very well the melting curve in
the covered range.

Our data are in excellent agreement with those reported by Bridgman\til\cite{Bri1914} where they overlap
($300<T<366\misura{K}$). Comparison with the work of Grace and Kennedy\til\cite{GraKen1967} also shows a good
agreement up to 373$\misura{K}$ and 1.27$\misura{GPa}$ but their melting points start to deviate from ours at
higher temperatures, reaching a pressure difference of 0.45$\misura{GPa}$ at 423$\misura{K}$. This difference is
outside our experimental uncertainty. We also note that a difference of the same magnitude was observed between
the argon melting points measured by Grace and Kennedy and data from other authors (see
Ref.~[\onlinecite{DatLou2000}] and references therein).

At higher temperature, the only melting point to compare our data with is the one reported by Iota and
Yoo\cite{IotYoo2001} at 4~GPa and 640~K\til.  We find here that this point is well inside the fluid domain and
overestimates the melting temperature by about 100~K at 4~GPa or underestimates the melting pressure by 2~GPa at
640~K. This discrepancy could be explained by a temperature overestimation in Ref.~[\onlinecite{IotYoo2001}]; as
a matter of fact, since the authors used ruby luminescence for pressure measurements, which has a 0.0073(1)~nm/K
temperature dependence\til\cite{DatLeT1997}, a temperature overestimation of 45~K would lead to a 0.9~GPa
pressure underestimation, which brings their point in nice agreement with ours.

The melting curve of molecular compounds are usually well described by semi-empirical melting laws like the
Simon-Glatzel (S-G)\til\cite{SG1929} or Kechin\til\cite{Kec2001} equations. The first one can be written as:
\begin{equation}
T = T_0 \biggl(1+\frac{P-P_0}{a}\biggr)^\frac{1}{b} \label{SG}
\end{equation}
where $T_0=216.59$~K and $P_0=0.518$~MPa are the triple point temperature and pressure respectively, and $a$ and
$b$ are two fit parameters. Fitting this expression to our data gives $a = 0.403(5)\misura{GPa}$ and  $b =
2.58(1)$. Figure \ref{melting} shows that this form reproduces very well our data, with a rms deviation of
$3.7$~K. The Kechin equation was proposed as an improvement over the Simon-Glatzel one and is able to represent
melting curves with negative slopes or going through a maximum melting temperature; it can be expressed as:
\begin{equation}
T = T_0 \biggl(1+\frac{P-P_0}{a}\biggr)^\frac{1}{b}\exp[c(P-P_0)] \label{Kechin}
\end{equation}
In the case of \CO, the melting temperature is a regular increasing function of pressure and a fit to the Kechin
equation only gives a small improvement over the S-G law. The obtained parameters are: $a=0.443(9)\misura{GPa}$,
$b=2.41(3)$ and $c=4.7(7)\times 10^{-3}\misura{GPa^{-1}}$, with a rms deviation of 3.5~K.

In addition to the visual observation of melting, we used Raman spectroscopy in order to identify the solid phase
in equilibrium with its melt. Fig.~\ref{Ramanspectra} reports spectra collected along the melting curve at 430,
733 and 800 K after subtraction of the liquid diffusion background. Spectra of solid phase I at 300 K, 2.1 GPa
and of solid phase IV at 580 K, 15 GPa are added for comparison. Despite the line broadening at high
temperatures, the three phonon modes of phase I are clearly recognizable on the spectra of the melting solid up
to 800 K. The absence of solid-solid phase transition can also be inferred from the continuous evolution of the
melting curve. Solid I is thus the stable phase along the melting curve up to 800 K and 11.1 GPa at least. This
result contrasts with the phase diagram reproduced in Fig.~1, since the reported I-IV transition
line\cite{IotYoo2001} should intersect our melting line at ca. 700~K and 7.8~GPa. Our findings suggest that the
I--IV--Fluid triple point is  at higher temperature.

\begin{figure}
\includegraphics[width=8.5cm]{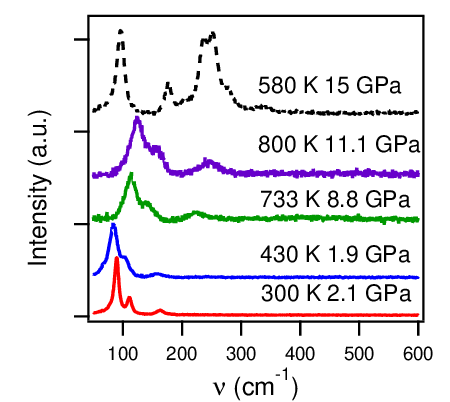}
\caption{}
\end{figure}

\subsection{Refractive index}

\begin{figure}
\centering
\includegraphics[width=8.5cm]{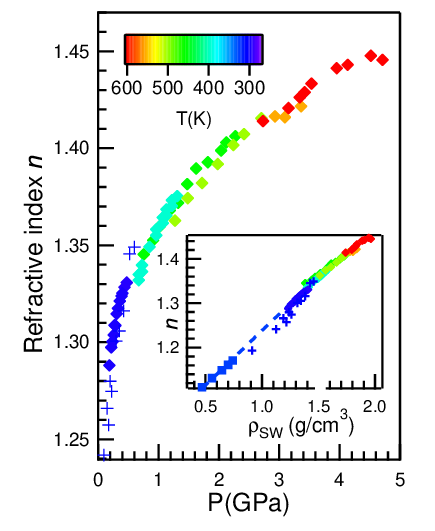}
\caption{}
\label{FigIndex}
\end{figure}

The refractive index $n$ of fluid \CO was measured along several isotherms (298, 395, 454, 500, 550 and 599K)
between 0.2 and 4.7~GPa, on two different samples. The evolution of $n$ with pressure and temperature is
represented in Fig.~\ref{FigIndex}, together with literature values available in the same pressure
range\til\cite{ShiKit1993}. In the inset, the same data are shown with respect to the density $\rho_{SW}$
calculated with the SWEOS. It is apparent from the latter plot that, within error bars, the refractive index does
not exhibit an intrinsic temperature dependence, but depends only on the density in the range of our experiments.
As a matter of fact, since the refractive index is a property related to the electronic structure of the
material, it is expected that density variations will have a greater influence than temperature variations, unless the
material undergoes temperature-driven structural changes.

The following equation was found to reproduce our refractive index data together with those measured in the
critical region by Sengers \textit{et al.}\til\cite{SenStr1971}:

\begin{equation}\label{eq:index1}
 n(\rho_{SW})=1 + 0.2153 \rho_{SW}+0.03868 \rho_{SW}^2-0.01563 \rho_{SW}^3
\end{equation}

The data from Ref.~\onlinecite{ShiKit1993} were not used in the fit because of their large scatter and poor
agreement with our measurements. Eq.\til(\ref{eq:index1}) was used here only as an intermediate step to determine
values of the refractive index at the $P-T$ conditions of the Brillouin measurements. As will be shown below, the
SWEOS is found to underestimate the density of fluid \CO at HP-HT and another expression of $n(\rho)$ will be
given with respect to our new equation of state in section 4.

\subsection{Brillouin scattering}

\begin{figure}
\centering
\includegraphics[width=8.5cm]{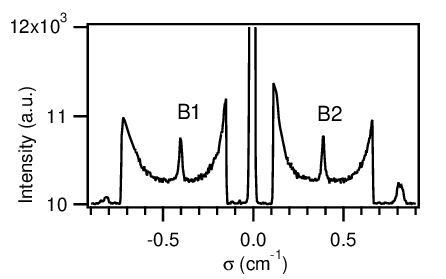}
\caption{}
\label{Brillexp}
\end{figure}

We performed Brillouin scattering experiments on the fluid along five isotherms at 300, 400, 500, 600 and
700$\misura{K}$ up to the respective freezing pressures. The Brillouin peaks were easily observable up to our
highest temperature and pressure values ($700\misura{K},8\misura{GPa}$): a typical Brillouin spectrum collected
in these conditions is shown in Fig.\til\ref{Brillexp}.

\begin{figure}
\centering
\includegraphics[width=8.5cm]{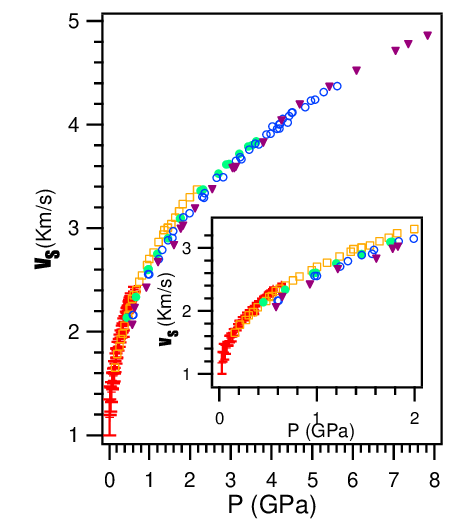}
\caption{}
\label{velexp}
\end{figure}

The sound velocity is deduced from the Brillouin frequency shift using Eqs.\til(\ref{BrillouinShift}) and
(\ref{eq:index1}). We note that, although our index measurements only extends to 599 K and 4.7 GPa, the
insensitivity of $n$ with respect to temperature changes and its smooth variation with respect to density gives
good confidence in extrapolating Eq.~(\ref{eq:index1}) in the near region. The evolution of $v_s$ with pressure
and temperature is represented in Fig.\til\ref{velexp}. The temperature
dependence of the sound velocity is small compared to its pressure dependence and becomes  almost unobservable above 4~GPa.

\begin{figure}
\centering
\includegraphics[width=8.5cm]{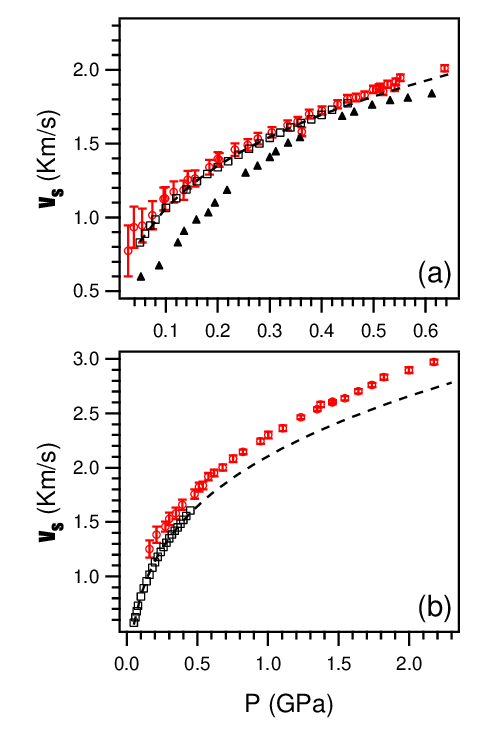}
\caption{}
\label{cfrShi_Pit}
\end{figure}

There are only two previous sources of experimental data for the high-pressure sound velocity which can be
compared to ours; the first one is from Pitaevskaya and Bilevich\til\cite{PitBil1973} who performed ultrasound
measurements up to 0.45~GPa and 473~K, with an estimated uncertainty of 1-2\%; the second one is from Shimizu
\textit{et al.}\til\cite{ShiKit1993}, who performed Brillouin scattering experiments at room temperature up to
0.55~GPa. Fig.\til\ref{cfrShi_Pit} shows the comparison at 300 K (top) and 400 K (bottom) of the experimental
data, together with the predictions from the SWEOS. At both temperatures, the sound velocities determined from
the SWEOS reproduce very well the ultrasonic measurements of Ref.~[\onlinecite{PitBil1973}]: there is a maximum
deviation of 1\% at 300K and 2\% at 400 K, i.e. within the experimental error bars. At 300~K, Shimizu \emph{et
al.}'s sound velocities are smaller than those of Pitaevskaya and Bilevich with differences up to 40\%. Our
results are in much better agreement with those of Pitaevskaya and Bilevich, although there is a systematic
deviation towards higher values of about 3\%. At 400~K, our sound velocities are larger than the ultrasonic ones
by about 8\% up to 0.45 GPa, and the deviation with respect to the SWEOS increases up to 9\% at 2.17 GPa.

\begin{figure}
\centering
\includegraphics[width=8.5cm]{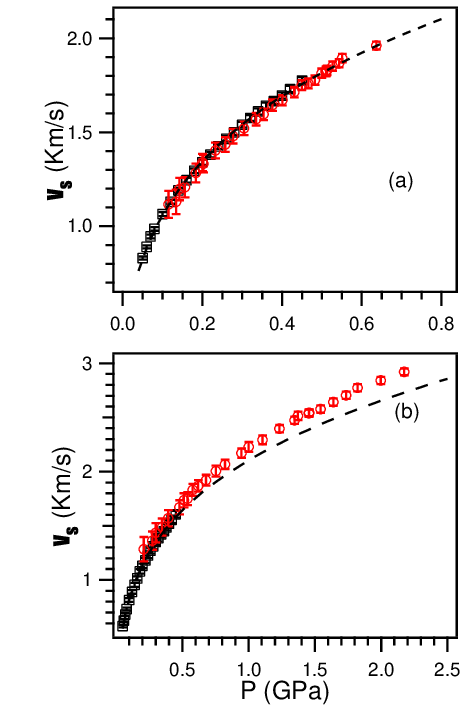}
\caption{}
\label{velcorr}
\end{figure}

Since the frequencies at which the ultrasonic and Brillouin experiments probe the sound velocity are different,
the apparent discrepancy between the two data sets suggests the presence of a dispersion effect as the one
observed in a relaxation phenomenon. As a matter of fact, in a molecular fluid like \CO, a thermal relaxation of
the vibrational degrees of freedom of the molecule is expected\cite{PhysicalAcoustics} whenever the relaxation
time $\tau_v$ of these vibrations, i.e. the time needed to reach thermal equilibrium between vibrational and
translational modes, becomes smaller than the probe period $1/\omega$. As a consequence, for $\omega\tau_v\sim 1$
the sound velocity goes from its low-frequency (``static") value $v_s^0$ to an unrelaxed value $v_s^\infty$.

The presence of a thermal relaxation in gaseous and liquid \CO is  a long-known and well documented
phenomenon\til\cite{Pielemeier1939,HenPes1957,MadLit1961}. The relaxation frequency $\omega_r$ was found to be a
linear function of density in the gas phase, whereas it increases more rapidly in the liquid. The values for
$\omega_r$ lie in the range 0.016-25$\misura{MHz}$ for pressures between 0.7$\cdot 10^{-3}$ and
0.16$\misura{GPa}$. The ultrasonic measurements of Pitaevskaya and Bilevich were performed at frequencies between
0.36 and 5$\misura{MHz}$
 in the pressure range 0.05--0.45$\misura{GPa}$; this is below the relaxation frequency, thus the observed sound velocities correspond to $v_s^0$. On the other hand, in our Brillouin scattering experiments the frequency of the probed thermal sound waves was in the range of 4.6--26$\misura{GHz}$, which is much higher than the reported values of $\omega_r$.  We thus expect to be in the unrelaxed region, as already found in previous Brillouin studies at low pressure\til\cite{GamSwi1967,LaoSch1976}.

In order to test this hypothesis, we calculated the predicted thermodynamic values of the sound velocity obtained
by applying the procedure described in Appendix A to our measured Brillouin velocities. They are compared to
ultrasonic and SWEOS velocities in Fig.\til\ref{velcorr}. It can be seen that at room temperature the agreement
with ultrasonic measurements is now very good, thus supporting the fact that Brillouin frequencies are higher
than the relaxation frequency. At 400$\misura{K}$ our "corrected" sound velocities remain about 4\% larger than
the ultrasonic ones up to 0.45 GPa, but they now both agree within error bars. At higher pressures, the deviation
from SWEOS predictions becomes larger than the experimental uncertainty, which shows that this EoS becomes
increasingly inaccurate outside its validity range. At 700~K and 8~GPa, the SWEOS underestimates the sound
velocity by about 9\%.

For the purpose of integration, as described below, we looked for a convenient analytical form to represent both
pressure and temperature dependence of the Brillouin sound velocity. We found that the following relation
reproduces well our data, except for a few points at low pressure:
\begin{equation}
\ln(v_s)=(a_{0}+a_{1}T)+(b_{0}+b_{1}T)\ln(P) \label{eq:lnv_vs_lnP}
\end{equation}
with $a_0=0.926(5)$, $a_1=-0.00026(1)$, $b_0=0.279(3)$ and $b_1=0.000134(7)$ for $v_s$ in km/s, $P$ in GPa and T in
K (numbers in parentheses give the standard error on the last digit).

\section{Calculation of the equation of state}

From the sound velocity data, corrected for the thermal relaxation, the equation of state of fluid \CO can be
obtained by recursively integrating the following equations:
\begin{equation}\label{EOS1}
\frac{\partial \rho}{\partial P}\Biggl|_T = \frac{1}{(v_s^0)^2}+\frac{T \alpha^2}{C_p}
\end{equation}

\begin{equation}
\frac{\partial c_p}{\partial P}\Biggl|_T = -T \frac{\partial^2 V}{\partial T^2}\Biggl|_P \label{EOS2}
\end{equation}
where $\rho,\; V,\; \alpha,\;$ and $c_p$ are the density, specific volume, volume thermal expansion and isobaric
heat capacity respectively. The method consists in integrating Eq.~(\ref{EOS1}) in a small temperature range
around the desired isotherm, by first neglecting the second term giving the correction between isothermal and
adiabatic compressions. From the value of $v_s^\infty$ given by Eq.~(\ref{eq:lnv_vs_lnP}), $v_s^0$ is calculated
using Eqs.~(\ref{corrv})-(\ref{PlanckEinstein}). The second term of Eq.~(\ref{EOS1}) is then evaluated using the
deduced value for $\alpha =1/V(\partial V/\partial T)$ and the one of $c_p$ obtained by integrating
Eq.\til(\ref{EOS2}). Eq.\til(\ref{EOS1}) is integrated again and the procedure iterated until convergence.  This
procedure was first tested using the sound velocity derived from the SWEOS (which directly gives $v_s^0$); the
calculated $\rho$, $\alpha$ and $c_p$ were in excellent agreement with those directly derived from the SWEOS.

The integration starts at a point $P_0$ for which density, $\alpha$, $c_p$ and $c_v$ need to be known. We chose
$P_0=0.25$~GPa, value above which Eq.~(\ref{eq:lnv_vs_lnP}) provides a reliable representation of our data at all
temperatures. The starting values for the thermodynamic quantities were calculated using the SWEOS, which
provides a consistent frame for the evaluation of any thermodynamic property. The uncertainty at this pressure
estimated by Span and Wagner is $\pm1$\% for density and $\pm2$\% for $c_p$.

\begin{figure}
\centering
\includegraphics[width=8.5cm]{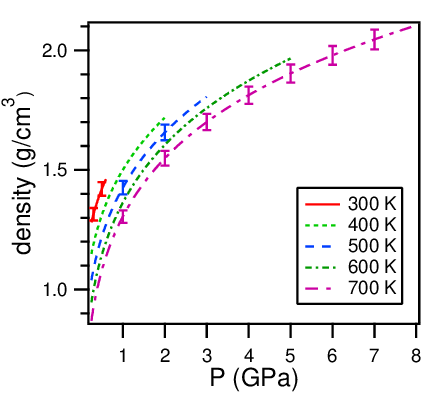}
\caption{}
 \label{rhograph}
\end{figure}

The calculated density vs pressure curves for \mbox{$300<T<700$~K} are shown in Fig.\til\ref{rhograph}. The
propagation of experimental error bars $\Delta v_s$ on $v_s$ to the calculated $\rho$ was estimated by integrating Eqq.~\ref{EOS1} and~\ref{EOS2} using $v_s \pm \Delta v_s$. 
The resulting error  $\Delta \rho / \rho$  reaches 0.6\%. However, this estimation does not include the
uncertainty of 1\% of the initial density value given by the SWEOS. Furthermore, the approximations underlying
the procedure applied to correct for the thermal relaxation introduce a larger uncertainty on the sound velocity.
The difference between the density calculated with and without thermal correction on the velocity data is at most
2\%. We thus indicate 2\% as a reasonable estimate of the error for the present density data.

When expressed on a logarithmic scale for $\rho$ and $P$, an analytical relation was found that well reproduces
our densities in the covered $P-T$ range:
\begin{equation}
\ln(\rho)=\sum_{i=0}^{2}\sum_{j=0}^3 a_{ij}T^{i}(\ln(P))^j \label{EOSfluid}
\end{equation}
with the density expressed in g/cm$^3$, $P$ in GPa and $T$ in K. The  $a_{ij}$ parameters, obtained by a
least-squares fit, are listed in Table\til\ref{fitEOS}.

The evolution of  $n$ with density, according to our new equation of state, is well described by the polynomial
form:
\begin{equation}
n(\rho)=1+0.21(1)\rho+0.04(1)\rho^2-0.017(5)\rho^3
\end{equation}

Fitting Eq.~(\ref{eq:lnv_vs_lnP}) to the thermodynamic values of the sound velocity deduced from our calculations
leads to the following coefficients: $a_0=0.9249(5)$, $a_1=-0.000392(1)$, $b_0=0.2683(6)$ and $b_1=0.000197(1)$.

As mentioned in the Introduction, there are a number of phenomenological equations of state for fluid \CO. These
equations have usually been parameterized in order to accurately reproduce the properties at low pressure and
around the critical point; their predictions at high pressure need to be tested against experimental data.

\begin{figure}
\centering
\includegraphics[width=8.5cm]{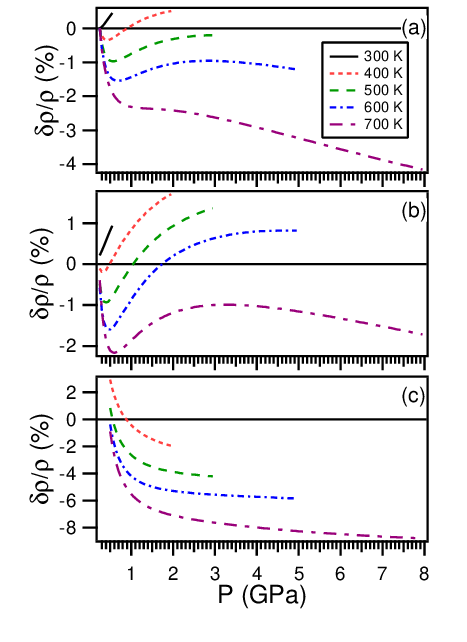}
\caption{}
\label{devrho}
\end{figure}

In Fig.\til\ref{devrho}\til(a) and (b) we compare our calculated EoS with the SWEOS and the PSEOS, respectively.
As it can be seen, our results show that the fluid becomes less compressible than predicted by the SWEOS above 500~K, with a difference in density which grows continuously  larger with temperature, up to -4.2\% at 8 GPa and 700 K. The PSEOS densities remain within
about 2.2\% of our data and the general agreement is better than for the SWEOS, although this EoS also
tends to overestimate the density at HP-HT.

Comparison with the classical molecular dynamics calculations of Belonoshko and Saxena\til\cite{BelSax1991} is
reported in Fig.\til\ref{devrho}\til(c). For these simulations, an effective intermolecular pair potential of the
exp-6 form was used. A fit to the densities obtained at various $P$-$T$ points was provided by the authors for
$0.5<P<100$~GPa and $400<T<4000$~K. The discrepancy with respect to the presently determined densities is within
2\% at 400 K but increases with temperature up to about -8.8\% at 8 GPa and 700 K, largely outside the present
uncertainty. This finding indicates
that the effective potential used in these simulations is too soft at high P
and T; this could be corrected by adopting a larger value for the stiffness parameter $\alpha$ of the exp-6
potential.

\squeezetable
\begin{table}
        \caption{}
                         \label{fitEOS}
             \centering
        \begin{ruledtabular}
        \begin{tabular}{ccccc}

          $_i\diagdown^j$ & 0 & 1 & 2 & 3\\
         \hline
                     $0$ & $0.6521(5)$ & $0.0301(7)$ & $-0.0139(8)$ & $-0.0150(6)$\\
                     $1$ & $-0.000700(2)$ & $0.000520(3)$ & $8.1(3)\cdot10^{-5}$ & $4.0(2)\cdot10^{-5}$\\
                     $2$ & $2.14(2)\cdot10^{-7}$ & $-2.75(2)\cdot10^{-7}$ &$-1.28(3)\cdot10^{-7}$ & $-2.1(2)\cdot10^{-8}$  \\
         \end{tabular}
        \end{ruledtabular}

\end{table}

\section{Conclusions}

In this work we have first accurately determined the melting curve of \CO up to 800 K by visual observation of
the solid-fluid equilibrium and  \textit{in-situ} $P$-$T$ measurements. The melting line shows a continuous,
monotonic increasing evolution with pressure in this $P$-$T$  range, and is very well reproduced by melting laws
of the Simon-Glatzel or Kechin form. The present determination is in very good agreement with the work of
Bridgman\til\cite{Bri1914} up to 366 K, but significantly deviate at higher $T$ from data of other
authors\cite{GraKen1967, IotYoo2001}. Furthermore, we found that the solid phase in equilibrium with the fluid
remains phase I up to $800\misura{K}$, indicating that the I-IV-Fluid triple point is located at higher
temperature than previously thought.

Secondly, we have reported the first measurements of the refractive index and Brillouin frequency shift in the
temperature and pressure range [300--700]$\misura{K}$ and [0.1--8]$\misura{GPa}$. From the deduced sound
velocities, corrected for the effect of thermal relaxation, we have calculated the equation of state of fluid \CO
with an estimated uncertainty of $\Delta \rho =2$\% (Eq.~\ref{EOSfluid}, table~\ref{fitEOS}), and compared it with various EoS proposed in the literature. The model of
Pitzer and Sterner provides the best agreement at the highest P-T conditions, but all tested models were found to
overestimate the compressibility of the fluid at high temperature. This work should thus be useful in order to
better constrain these EoS in the high pressure regime.

\appendix

\section{Thermal relaxation and dispersion of the sound velocity.}
The thermodynamic definition of the adiabatic sound velocity $v_s^0$ is:
\begin{equation}
(v_s^0)^2=(c_p^0/c_v^0)\left.\frac{dP}{d\rho}\right|_T \label{appeq1}
\end{equation}
where $c_p^0$ and $c_v^0$ are the thermodynamic (or``static") values of the specific heats at constant pressure
and constant volume respectively. As long as the frequency $\omega$ at which the sound waves are probed remain
below any relaxation frequency, then the measured sound velocity is given by Eq.~(\ref{appeq1}).

The specific heats include several contributions, one of which, $c_{vib}$, is due to the vibrational degrees of
freedom of the molecules. In the condition $\omega\tau_v\simeq 1$ ($\tau_v$ being the thermal relaxation time),
the contribution to the specific heat from the vibrational modes is modulated by their degree of relaxation. For
$\omega\tau_v\gg 1$, these modes will be frozen-like and their contribution to the specific heats relaxes out
\cite{PhysicalAcoustics}, so that the effective specific heats will be $c_{i}^{\infty}=c_{i}^0-c_{vib}$, with
$i=p,v$. The unrelaxed sound velocity may then be calculated as:
\begin{equation}
(v_\infty)^2=(c_p^\infty/c_v^\infty)\left.\frac{dP}{d\rho}\right|_T
\end{equation}
and thus:
\begin{equation}
(v_0/v_\infty)^2=(c_p^0/c_v^0)(c_v^\infty/c_p^\infty) \label{corrv}
\end{equation}
The value of $c_{vib}$ may be deduced from the knowledge of the vibrational frequencies $\nu_i$ and their
degeneracies $g_i$, using the Planck-Einstein formula ($R$ is the ideal gas constant):

\begin{equation}
c_{vib}=R \sum_i \frac{g_i  (h\nu_i)^2/k_BT^2 \exp(-h\nu_i/k_BT)} {(1-exp(-h\nu_i/k_BT))^2} \label{PlanckEinstein}
\end{equation}

To our knowledge, there are no data on the vibrational frequencies of the fluid in our $P-T$ range; however, the
measurements made on solid I showed that these modes are very weakly pressure and temperature dependent
\til\cite{HanJon1981,LuHof1995,GioGor2006}, which should also be valid for the fluid. The estimation of $c_{vib}$
was thus based on the gas values for the vibrational frequencies\cite{Pielemeier1939}, and assumed to be pressure
independent.

\bibliographystyle{apsrev}
\bibliography{biblio_CO2}

\newpage
Figures\\

\begin{itemize}

\item Fig.~\ref{phasedia}: (Color online) Phase diagram of \CO as presently known up to 20~GPa from the work of
Ref.\til[\onlinecite{IotYoo2001}], and modified to take into account the experimental results of
Ref.~[\onlinecite{GorGio2004}]. The dashed line is the kinetic barrier between phases III and II. The symbols
show the experimental melting points available in the literature. $\circ$: Bridgman\til\cite{Bri1914}; $\times$:
Grace and Kennedy \til\cite{GraKen1967}; $\blacksquare$: Iota and Yoo \til\cite{IotYoo2001}. \\

\item Fig.~\ref{melting}: (Color online) Experimental melting points determined from three separate experiments
($\bullet, \blacktriangledown, \blacktriangle$) along with literature data ($\circ$ from Bridgman\cite{Bri1914},
$\square$ from Grace and Kennedy\cite{GraKen1967} and $\blacksquare$ from Iota and Yoo\cite{IotYoo2001}).  The
solid line is the fit to our data using the the Simon-Glatzel melting law [Eq.~(\ref{SG}) with
a=$(0.403\pm0.005)~\rm{GPa}$, b=$(2.58\pm0.01)$]; the dotted lines are guides to the eyes. The inset shows a
photograph of the single crystal (a) in equilibrium with its melt (b) taken
at 560~K.  The pressure gauges are visible: ruby (c), cBN (d) and \samarion (e).\\

\item Fig.~\ref{Ramanspectra}: Raman spectra collected on the solid phase in equilibrium with the melt along the
melting curve at 430 K, 733 K and 800 K. A spectrum of phase I at 300~K and 2.1~GPa and one of phase IV at 580~K
and 15~GPa (dashed line) are shown for comparison. \\

\item Fig.~\ref{FigIndex}: (Color online) Refractive index $n$ of fluid \CO as a function of pressure. Diamonds:
this work; crosses: Ref.~[\onlinecite{ShiKit1993}]; squares: Ref.~[\onlinecite{SenStr1971}]. Inset: same data
plotted as a function of the SWEOS predicted density $\rho_{SW}$. The dashed line is the fit to our data and
those of Ref.~[\onlinecite{SenStr1971}] [Eq.~(\ref{eq:index1})].\\

\item Fig.~\ref{Brillexp}: Brillouin spectrum from the \CO sample collected at 700$\misura{K}$ and
8$\misura{GPa}$. The intensity was scaled to emphasize the two Brillouin peaks B1 and B2 (Stokes and anti-Stokes
components). The weak features at $\sim\pm0.8\misura{cm^{-1}}$ are the "ghosts" of the neighboring interference
orders resulting from the tandem arrangement of the Sandercock interferometer \cite{Sander}. \\

\item Fig.~\ref{velexp}: (Color online) Evolution of the sound velocity as determined by the present Brillouin
scattering and refractive index measurements, along isotherms at 300($+$), 400($\square$), 500($\bullet$),
600($\circ$) and 700$\misura{K}$($\blacktriangledown$). The error bars on sound velocities include the ones for
pressure and are more important at low pressure due to the larger sensitivity of the sound velocity to variations
of pressure. The inset shows a zoom of the 0-2~GPa region. \\
\item Fig.~\ref{cfrShi_Pit}: Comparison between present sound velocity measurements (circles) at 300 K (a) and
400 K (b) with the available literature data. Squares: Pitaevskaya and Bilevich\cite{PitBil1973}; triangles:
Shimizu et al\cite{ShiKit1993}. The sound velocities predicted from the SWEOS are reported as dashed lines. \\

\item Fig.~\ref{velcorr}: Comparison between the thermodynamic sound velocity obtained by "correcting" the
Brillouin velocities for the effect of thermal relaxation ($\circ$), with the ultrasonic ($\square$) and SWEOS
(dashed line) velocities at 300~K (a) and 400$\misura{K}$\til(b). \\

\item Fig.~\ref{rhograph}: Calculated densities along several isotherms in the range 300-700 K. \\

\item  Fig.~\ref{devrho} (Color online) The relative deviation $(\rho_{exp}-\rho_{eos})/\rho_{exp}$ of our
calculated densities with respect to the predictions from SWEOS\cite{SpaWag1996}\til(a),
PSEOS,\cite{PitSte1994}\til(b) and the simulations of Belonoshko and Saxena\til\cite{BelSax1991}\til(c) is
reported for different isotherms between 300 and 700~K. \\

\end{itemize}

Tables \\

\begin{itemize}
\item Tab.~\ref{fitEOS}: Values of the parameters $a_{ij}$ obtained by fitting Eq.\til(\ref{EOSfluid}) to the calculated densities.\\
\end{itemize}
\end{document}